\documentclass[aps,prd,onecolumn,groupedaddress,showpacs,nofootinbib,amssymb]{revtex4}
\usepackage{graphicx,bm,color}
\usepackage{amsmath}
\usepackage{amssymb}
\usepackage{amsfonts}

\newcommand{\be}{\begin{equation}}
\newcommand{\ee}{\end{equation}}
\newcommand{\bea}{\begin{eqnarray}}
\newcommand{\eea}{\end{eqnarray}}
\newcommand{\beaa}{\begin{eqnarray*}}
\newcommand{\eeaa}{\end{eqnarray*}}

\newcommand{\nn}{\nonumber \\}
\newcommand{\e}{\mathrm{e}}
\newcommand{\tr}{\mathrm{tr}\,}

\allowdisplaybreaks[4]

\begin{document}

\title{Cosmology and Stability in Scalar-Tensor bigravity}

\author{Kazuharu Bamba$^1$, Yusuke Kokusho$^2$, Shin'ichi Nojiri$^{1,2}$ 
and Norihito Shirai$^2$}

\affiliation{
$^1$ Kobayashi-Maskawa Institute for the Origin of Particles and
the Universe, Nagoya University, Nagoya 464-8602, Japan \\
$^2$ Department of Physics, Nagoya University, Nagoya
464-8602, Japan \\
}

\begin{abstract}

The bigravity models coupled with two scalar fields are constructed. 
We show that a wide class of the expansion history of the 
universe, especially corresponding to dark energy and/or inflation, 
can be described by a solution of the bigravity model. 
We discuss the stability of the solution and give the conditions 
for the stability. 
We also explicitly construct a model which gives a stable solution.  
By using the stable model, for an arbitrary evolution of 
the universe expansion, we construct the Brans-Dicke like model 
which reproduces the evolution. 

\end{abstract}


\maketitle

\section{Introduction \label{Sec1}}

Although free massive gravity was established about 
seventy-five years ago in Ref.~\cite{Fierz:1939ix} (for a recent 
review, see \cite{Hinterbichler:2011tt}), it is difficult to 
obtain consistent interacting or non-linear models for a long time because 
such a model contains the Boulware-Deser ghost \cite{Boulware:1974sr,Boulware:1973my} 
in general and there also appears the van Dam-Veltman-Zakharov (vDVZ) 
discontinuity \cite{vanDam:1970vg} in the massless limit, $m\to 0$.
It is known that the discontinuity can be screened 
by the Vainshtein mechanism \cite{Vainshtein:1972sx} 
(see, for example, Ref.~\cite{Luty:2003vm}). 
%%%%%%%%%%%
Note that the extra degrees of freedom are cancelled by the ghost degrees of 
freedom. 
%%%%%%%%%%%

Recently, however, the study of the non-linear massive gravity has  
progressed remarkably and the ghost-free models, which are called the 
de Rham, Gabadadze, Tolley (dRGT) models, have been constructed and 
found for non-dynamical background metric 
in \cite{deRham:2010ik,deRham:2010kj,Hassan:2011hr} and for dynamical 
metric \cite{Hassan:2011zd}. 
The so-called minimal model first appeared in \cite{Hassan:2011vm}
(for the general proof of absence of ghost in massive gravity, 
see Ref.~\cite{Hassan:2011tf}). 
%%%%%%
Even in the dRGT models, the Vainshtein mechanism works 
although there does not appear ghost.
%%%%%%%
Since the models for the dynamical background metric contain 
two metric (or symmetric tensor fields), such models are called as 
bigravity models. 

The massive gravity models have been applied to cosmology 
in Refs.~\cite{Kluson:2012zz,Kluson:2012wf,Hassan:2011ea,D'Amico:2011jj}. 
For the bimetric gravity, some cosmological solutions including the ones 
describing the accelerating universe have been investigated 
\cite{Damour:2002wu,Volkov:2011an,vonStrauss:2011mq,Berg:2012kn,Nojiri:2012zu,Nojiri:2012re,Bamba:2013fha,AKMS-TSK}. 
In case of the $dRGT$ model, it became clear that 
the flat Friedmann-Robertson-Walker (FRW) cosmology is absent \cite{D'Amico:2011jj}. 
In this paper, we consider the bigravity models with scalar fields. 
We show that there are models which admit the stable solution describing 
the FRW solution with the spatially flat metric. 

The organization of the paper is as follows. 
In the next Section, we show one of the difficulties in massive gravity theories 
with a non-dynamical background metric to construct a model 
in which the non-trivial FRW cosmology is realized. 
Then in Section \ref{Sec3}, we consider bigravity models where the background 
metric is also dynamical. We construct bigravity models coupled 
with two scalar fields and show that a wide class of the expansion history 
of the universe can be described by a solution of the bigravity model. 
The solution is, however, not always stable under the perturbation, 
that is, the small perturbation to the solution grows up in general. 
In Section \ref{Sec4}, we investigate the stability of the solutions and 
give the conditions for the stability. Furthermore we explicitly build 
a model which gives a stable solution. 
In Section \ref{Sec5}, by using the stable model obtained 
in Section \ref{Sec4}, 
for an arbitrary evolution of the universe expansion, 
we construct the Brans-Dicke like model which reproduces the evolution. 
The last Section is devoted to conclusions.

\section{Difficulties of cosmology by massive gravity with scalar field \label{Sec2}}

In this section, we show that it is very difficult to construct a massive 
gravity model coupled with a scalar field. Such a coupling could lead to 
a solution describing the FRW space-time with the vanishing spatial curvature. 

The starting action is given by
\be
\label{massivegravity}
S_\mathrm{mg} = M_g^2\int d^4x\sqrt{-\det g}\,R^{(g)}
+2m^2 M_\mathrm{eff}^2 \int d^4x\sqrt{-\det g}\sum_{n=0}^{4} \beta_n\,
e_n \left(\sqrt{g^{-1} f} \right) \, ,
\ee
where $R^{(g)}$ is the scalar curvature for $g_{\mu \nu}$ and
$f_{\mu \nu}$ is a non-dynamical reference metric.
The tensor $\sqrt{g^{-1} f}$ is defined by the square root of
$g^{\mu\rho} f_{\rho\nu}$, namely, 
$\left(\sqrt{g^{-1} f}\right)^\mu_{\ \rho} \left(\sqrt{g^{-1}
f}\right)^\rho_{\ \nu} = g^{\mu\rho} f_{\rho\nu}$.
For general tensor $X^\mu_{\ \nu}$, $e_n(X)$'s are defined by
\begin{align}
\label{ek}
& e_0(X)= 1  \, , \quad
e_1(X)= [X]  \, , \quad
e_2(X)= \tfrac{1}{2}([X]^2-[X^2])\, ,\nn
& e_3(X)= \tfrac{1}{6}([X]^3-3[X][X^2]+2[X^3])
\, ,\nn
& e_4(X) =\tfrac{1}{24}([X]^4-6[X]^2[X^2]+3[X^2]^2
+8[X][X^3]-6[X^4])\, ,\nn
& e_k(X) = 0 ~~\mbox{for}~~k>4 \, .
\end{align}
Here $[X]$ expresses the trace of arbitrary tensor
$X^\mu_{\ \nu}$: $[X]=X^\mu_{\ \mu}$.

We add the following terms to the action (\ref{massivegravity}):
\be
\label{mg1}
S_\phi = - M_g^2 \int d^4 x \sqrt{-\det g}
\left\{ %\frac{3}{2} 
\frac{1}{2}
%%%%%
g^{\mu\nu} \partial_\mu \phi \partial_\nu \phi
+ V(\phi) \right\} 
%+ \int d^4 x \mathcal{L}_\mathrm{matter}
%\left( \e^{\varphi} g_{\mu\nu}, \Phi_i \right)
\, .
\ee
By the conformal transformation 
$g_{\mu\nu} \to \e^{-\varphi(\phi)} g^{\mathrm{J}}_{\mu\nu}$,
the total action $S_\mathrm{BD} = S_\mathrm{mg} + S_\varphi$
is transformed as
\begin{align}
\label{mg2}
S_\mathrm{BD} =& + M_g^2 \int d^4 x \sqrt{-\det g^{\mathrm{J}}}
\left\{ \e^{-\varphi} R^{\mathrm{J}(g)} - %\frac{3}{2} 
\frac{1}{2} 
%%%%%%%%%%
\e^{-\varphi(\phi)} 
\left( 1 - 
%%%%
3
%%%%
\varphi'\left(\phi\right)^2 \right) 
g^{\mu\nu} \partial_\mu \phi \partial_\nu 
\phi - \e^{-2\varphi} V(\varphi) \right\}\nn
& + 2m^2 M_\mathrm{eff}^2 \int d^4x\sqrt{-\det g^{\mathrm{J}}}\sum_{n=0}^{4}
\beta_n \e^{\left(\frac{n}{2} -2 \right)\varphi} e_n
\left(\sqrt{{g^{\mathrm{J}}}^{-1} f} \right) 
%+ \int d^4 x \mathcal{L}_\mathrm{matter}
%\left( g^{\mathrm{J}}_{\mu\nu}, \Phi_i \right) 
\, .
\end{align}
Thus we obtain a Brans-Dicke type model. 
%%%%%
Then if we have a solution where the scalar field is not constant but depends on the time coordinate and the 
space-time is the arbitrary FRW background even if flat background, we  may obtain arbitrary history of the 
expansion by the conformal transformation. As we will see in (\ref{mg6}) later, however, the scalar field should be 
constant, which is the motivation why we consider the bigravity. 
%%%%%%%%%%%%%%%%%

In the following, just for simplicity, we only investigate the minimal case \cite{Hassan:2011vm}
\be
\label{mg3}
S_\mathrm{mg} = M_g^2\int d^4x\sqrt{-\det g}\,R^{(g)} 
+2m^2 M_\mathrm{eff}^2 \int d^4x\sqrt{-\det g} \left( 3 - \tr \sqrt{g^{-1} f}
+ \det \sqrt{g^{-1} f} \right)\, .
\ee
%%%%
In terms of $e_n$ in (\ref{ek}), we find 
$3 - \tr \sqrt{g^{-1} f} + \det \sqrt{g^{-1} f}
= 3 e_0 \left( \left(\sqrt{g^{-1} f}\right)^\mu_{\ \nu}\right)
 - e_1 \left( \left(\sqrt{g^{-1} f}\right)^\mu_{\ \nu}\right)  
+ e_4 \left( \left( \sqrt{g^{-1} f}\right)^\mu_{\ \nu}\right) $. 
The minimal case could be a simplest but non-trivial case and proposed in  \cite{Hassan:2011vm}. 
In the model, the interaction between two metrics $g_{\mu\nu}$ and $f_{\mu\nu}$ is only given by 
the trace of $\left(\sqrt{g^{-1} f}\right)^\mu_{\ \nu}$. 
When we consider non-minimal models, the calculations becomes rather complicated but 
the quantitative structure in the arguments in this paper could not be changed. 
%%%%%
In order to evaluate $\delta \sqrt{g^{-1} f}$, two matrices $M$ and
$N$, which satisfy the relation $M^2=N$, are taken.
Since $\delta M M + M \delta M = \delta N$, we have
\be
\label{Fbi7}
\tr \delta M = \frac{1}{2} \tr \left( M^{-1} \delta N \right)\, .
\ee
For a while, we examine the Einstein frame action (\ref{mg3}) with
(\ref{mg1}) but matter contribution is neglected. 
Therefore by the variation over $g_{\mu\nu}$, we acquire 
\begin{align}
\label{Fbi8}
0 =& M_g^2 \left( \frac{1}{2} g_{\mu\nu} R^{(g)} - R^{(g)}_{\mu\nu} \right)
+ m^2 M_\mathrm{eff}^2 \left\{ g_{\mu\nu} \left( 3 - \tr \sqrt{g^{-1} f}
\right)
+ \frac{1}{2} f_{\mu\rho} \left( \sqrt{ g^{-1} f } \right)^{-1\, \rho}_{\qquad 
\nu}
+ \frac{1}{2} f_{\nu\rho} \left( \sqrt{ g^{-1} f } \right)^{-1\, \rho}_{\qquad 
\mu} \right\} \nn
& + M_g^2 \left[ \frac{1}{2} \left( %\frac{3}{2} 
\frac{1}{3}
%%%%%%%%%%%
g^{\rho\sigma} \partial_\rho %\varphi 
\phi
%%%%%
\partial_\sigma %\varphi
\phi
%%%%%%%%%%
+ V (
%\varphi
\phi
%%%%%%%%%%%
) \right) g_{\mu\nu} - %\frac{3}{2}
\frac{1}{2}
%%%%
\partial_\mu %\varphi 
\phi
%%%%%%%%%%%
\partial_\nu 
%\varphi 
\phi \right] \, .
\end{align}
We should note that $\det \sqrt{g} \det \sqrt{g^{-1} f } \neq \det \sqrt{f}$ in general. 
The variation of the scalar field %$\varphi$ 
$\phi $
%%%%%%%%%%%%%
yields 
\be
\label{scalareq}
0 = - %3 
\Box_g \phi + V' (\phi) \, ,
\ee
where $\Box_g$ is the d'Alembertian with respect to the metric $g$.
By multiplying the covariant derivative $\nabla_g^\mu$ with respect to the 
metric $g$ by Eq.~(\ref{Fbi8}) and using the Bianchi identity 
$0=\nabla_g^\mu\left( \frac{1}{2} g_{\mu\nu} R^{(g)} - R^{(g)}_{\mu\nu} 
\right)$ and Eq.~(\ref{scalareq}), we have 
\be
\label{identity1}
0 = - g_{\mu\nu} \nabla_g^\mu \left( \tr \sqrt{g^{-1} f} \right)
+ \frac{1}{2} \nabla_g^\mu \left\{ f_{\mu\rho} \left( \sqrt{ g^{-1} f } 
\right)^{-1\, \rho}_{\qquad \nu}
+ f_{\nu\rho} \left( \sqrt{ g^{-1} f } \right)^{-1\, \rho}_{\qquad \mu} 
\right\} \, .
\ee
In case of the Einstein gravity, the conservation law of the energy-momentum 
tensor depends on the Einstein equation. It can be derived from the Bianchi 
identity. 
In case of massive gravity, however, the conservation laws of 
the energy-momentum tensor of the scalar fields are derived from the scalar field equations. 
These conservation laws are independent of the Einstein equation. 
The Bianchi identities present the equation (\ref{identity1}) 
independent of the Einstein equation.

We assume the FRW universe for the metric $g_{\mu\nu}$ and 
the flat Minkowski space-time for $f_{\mu\nu}$ and use the conformal 
time $t=\tau$ for the universe with metric $g_{\mu\nu}$: 
\be
\label{Fbi10}
ds_g^2 = \sum_{\mu,\nu=0}^3 g_{\mu\nu} dx^\mu dx^\nu
= a(\tau)^2 \left( - d\tau^2 + \sum_{i=1}^3 \left( dx^i \right)^2\right) \, ,\quad
ds_f^2 = \sum_{\mu,\nu=0}^3 f_{\mu\nu} dx^\mu dx^\nu
= - d\tau^2 + \sum_{i=1}^3 \left( dx^i \right)^2 \, .
\ee
%%%%%%%%%%
The physical meaning of the metric $f_{\mu\nu}$ has not been clear 
although there are several conjectures as in \cite{Deffayet:2012zc}. 
%%%%%%%%%% 
The longitudinal scalar modes in the metric play the roles when 
we consider the background as in (\ref{Fbi10}) but these 
modes do not propagate, which may 
be found by considering the local Lorentz frame. 
The propagating modes could be massless tensor (massless graviton) and 
the massive tensor (massive graviton) and any scalar mode does not propagate.
%%%%%%%%%%

The $(\tau,\tau)$ component of (\ref{Fbi8}) reads
\be
\label{Fbi11}
0 = - 3 M_g^2 H^2 - 3 m^2 M_\mathrm{eff}^2
\left( a^2 - a \right) + \left(
%\frac{3}{4}
\frac{1}{4}
%%%%
{\dot\phi}^2 + \frac{1}{2} V (\phi) a(\tau)^2 \right) M_g^2 \, ,
\ee
and from $(i,j)$ components we find
\be
\label{Fbi12}
0 = M_g^2 \left( 2 \dot H + H^2 \right)
+  3 m^2 M_\mathrm{eff}^2 \left( a^2 - a \right) + \left(
%\frac{3}{4} 
\frac{1}{2}
%%%%%%%%%%%
{\dot\phi}^2 - \frac{1}{2} V (\phi) a(\tau)^2 \right) M_g^2 \, , 
\ee
with $H=\dot a / a$, where the dot denotes the derivative with respect to 
$t$. From Eq.~(\ref{identity1}), we have the following equation: 
\be
\label{identity3}
\frac{\dot a}{a} = 0\, .
\ee
Hence $a$ should be a constant $a=a_0$. 
This means that the only consistent solution for $g_{\mu\nu}$ 
is the flat Minkowski space. 
Furthermore by using (\ref{Fbi11}) and (\ref{Fbi12}), we find
\be
\label{mg6}
\dot \phi =0\, ,\quad 0 = 3 m^2 M_\mathrm{eff}^2 
\left( a_0^2 - a_0 \right) - \frac{1}{2} V_0 a_0^2 M_g^2 \, .
\ee
Since $\phi$ is a constant, we cannot obtain the expanding universe. 

\section{Bigravity with two scalar fields and cosmology \label{Sec3}}

In the last section, we have observed difficulties to construct the model 
which generates the expanding universe. In this section, we build models 
of cosmology by using the bigravity with two scalar fields. 
The bimetric gravity proposed in Ref.~\cite{Hassan:2011zd} 
includes two metric tensors $g_{\mu\nu}$ and $f_{\mu\nu}$. 
In addition to the massless spin-two field, corresponding to graviton, 
it contains massive spin-two field although massive gravity models 
only have the massive spin-two field. The Boulware-Deser 
ghost \cite{Boulware:1974sr} does not appear in such a theory. 

\subsection{Bigravity models with scalar fields}

We add the term containing the scalar curvature $R^{(f)}$ given by 
$f_{\mu\nu}$ to the action (\ref{massivegravity}) as follows 
\begin{align}
\label{bimetric}
S_\mathrm{bi} =&M_g^2\int d^4x\sqrt{-\det g}\,R^{(g)}+M_f^2\int d^4x
\sqrt{-\det f}\,R^{(f)} \nonumber \\
&+2m^2 M_\mathrm{eff}^2 \int d^4x\sqrt{-\det g}\sum_{n=0}^{4} \beta_n\,
e_n \left(\sqrt{g^{-1} f} \right) \, .
\end{align}
Here $M_\mathrm{eff}$ is defined by
\be
\label{Meff}
\frac{1}{M_\mathrm{eff}^2} = \frac{1}{M_g^2} + \frac{1}{M_f^2}\, .
\ee
%%%%%%%%%%%
There is a conjecture that the two dynamical metric may 
correspond to manifolds with two metric \cite{Deffayet:2012zc}. 
%%%%%%%%%%%%%%%%%%

We also involve the following terms given by two scalar fields $\varphi$ and 
$\chi$ in the action (\ref{bimetric}):
\begin{align}
\label{Fbi1}
S_\varphi =& - M_g^2 \int d^4 x \sqrt{-\det g}
\left\{ %\frac{3}{2} 
\frac{1}{2}
%%%%%%%%
g^{\mu\nu} \partial_\mu \varphi \partial_\nu \varphi
+ V(\varphi) \right\} + \int d^4 x \mathcal{L}_\mathrm{matter}
\left( g_{\mu\nu}, \Phi_i \right)\, ,\\
\label{Fbi7b}
S_\xi =& - M_f^2 \int d^4 x \sqrt{-\det f}
\left\{ %\frac{3}{2} 
\frac{1}{2}
%%%%%%%%%%%%%%%
f^{\mu\nu} \partial_\mu \xi \partial_\nu \xi
+ U(\xi) \right\} \, .
\end{align}

For simplicity, we start from the minimal case again 
\begin{align}
\label{bimetric2}
S_\mathrm{bi} =&M_g^2\int d^4x\sqrt{-\det g}\,R^{(g)}+M_f^2\int d^4x
\sqrt{-\det f}\,R^{(f)} \nonumber \\
&+2m^2 M_\mathrm{eff}^2 \int d^4x\sqrt{-\det g} \left( 3 - \tr \sqrt{g^{-1} f}
+ \det \sqrt{g^{-1} f} \right)\, .
\end{align}
For a while, we neglect the contributions from matters. 
By the variation over $g_{\mu\nu}$, we again find (\ref{Fbi8}). 
On the other hand, through the variation over $f_{\mu\nu}$, we acquire
\begin{align}
\label{Fbi9}
0 =& M_f^2 \left( \frac{1}{2} f_{\mu\nu} R^{(f)} - R^{(f)}_{\mu\nu} \right) \nn
& + m^2 M_\mathrm{eff}^2 \sqrt{ \det \left(f^{-1}g\right) } 
\left \{ - \frac{1}{2}f_{\mu\rho} \left( \sqrt{g^{-1} f} 
\right)^{\rho}_{\ \nu} - \frac{1}{2}f_{\nu\rho} \left( \sqrt{g^{-1} f} \right)^{\rho}_{\ \mu} 
+ \det \left( \sqrt{g^{-1} f} \right) f_{\mu\nu} \right\} \nn
& + M_f^2 \left[ \frac{1}{2} \left( %\frac{3}{2} 
\frac{1}{3}
%%%%%%%%%%%%%%%%
f^{\rho\sigma} \partial_\rho
\xi \partial_\sigma \xi
+ U (\xi) \right) f_{\mu\nu} - \frac{3}{2} \partial_\mu \xi \partial_\nu \xi
\right] \, .
\end{align}
The variations of the scalar fields $\varphi$ and $\xi$ lead to 
\be
\label{scalareq2}
0 = - %3 
\Box_g \varphi + V' (\varphi) \, ,\quad
0 = - %3 
\Box_f \xi + U' (\xi) \, , 
\ee
corresponding to (\ref{scalareq}). 
Here $\Box_f$ is the d'Alembertian with respect to the metric $f$. From Eq.~(\ref{Fbi8}) 
and the Bianchi identity, again we acquire (\ref{identity1}). 
Similarly by using the covariant derivative $\nabla_f^\mu$ with respect to the 
metric $f$, from (\ref{Fbi9}), we have 
\be
\label{identity2}
0 = \nabla_f^\mu \left[ \sqrt{ \det \left(f^{-1}g\right) } 
\left\{ - \frac{1}{2}\left( \sqrt{g^{-1} f} \right)^{ -1 \nu}_{\ \ \ \ \ \sigma} 
g^{\sigma\mu}  - \frac{1}{2}\left( \sqrt{g^{-1} f} \right)^{ -1 \mu}_{\ \ \ \ \sigma} 
g^{\sigma\nu} + \det \left( \sqrt{g^{-1} f} \right) f^{\mu\nu} \right\} \right]\, .
\ee
The identities (\ref{identity1}) and (\ref{identity2}) impose strong constraints 
on the solutions. Especially we investigate the solutions describing the FRW universe in the next subsection. 

\subsection{Reconstruction of bigravity models}

We examine whether we can construct models describing the arbitrarily 
given evolution of the expansion in the universe. 

We take the FRW universes for the metric $g_{\mu\nu}$ as in (\ref{Fbi10}) 
and use the conformal time $t=\tau$. Moreover, instead of (\ref{Fbi10}), 
we suppose the form of the metric $f_{\mu\nu}$ as follows
\be
\label{Fbi10b}
ds_g^2 = \sum_{\mu,\nu=0}^3 g_{\mu\nu} dx^\mu dx^\nu
= a(\tau)^2 \left( - d\tau^2 + \sum_{i=1}^3 \left( dx^i \right)^2\right) \, ,\quad
ds_f^2 = \sum_{\mu,\nu=0}^3 f_{\mu\nu} dx^\mu dx^\nu
= - c(\tau)^2 d\tau^2 + b(\tau)^2 \sum_{i=1}^3 \left( dx^i \right)^2 \, .
\ee
%%%%%%%%%%%%%%
We should note the assumption in(\ref{Fbi10} could be most general form 
if we assume the spatial part of the space-time is uniform, homogeneous, and 
flat. The redefinition of the time-coordinate always gives the form of 
$ds_g^2$ but there does not any more freedom to choose $c(\tau)=1$ 
nor $c(\tau)=b(\tau)$. 
%%%%%%%%%%%%%%%%%%
In this case, from the $(\tau,\tau)$ component of (\ref{Fbi8}) we find
\be
\label{Fbi11b}
0 = - 3 M_g^2 H^2 - 3 m^2 M_\mathrm{eff}^2
\left( a^2 - ab \right) + \left( %\frac{3}{4} 
\frac{1}{4}
%%%%%%%%%%%%%%%%
{\dot\varphi}^2
+ \frac{1}{2} V (\varphi) a(\tau)^2 \right) M_g^2 \, ,
\ee
and $(i,j)$ components yield
\be
\label{Fbi12b}
0 = M_g^2 \left( 2 \dot H + H^2 \right)
+  m^2 M_\mathrm{eff}^2 \left( 3a^2 - 2ab - ac \right) + \left( %\frac{3}{4} 
\frac{1}{4} {\dot\varphi}^2 - \frac{1}{2} V (\varphi) a(\tau)^2 \right) M_g^2 \, .
\ee
On the other hand, the $(\tau,\tau)$ component of (\ref{Fbi9}) leads to 
\be
\label{Fbi13}
0 = - 3 M_f^2 K^2 +  m^2 M_\mathrm{eff}^2 c^2
\left ( 1 - \frac{a^3}{b^3} \right )
+ \left( %\frac{3}{4} 
\frac{1}{4}{\dot\xi}^2 - \frac{1}{2} U (\xi) 
c(\tau)^2 \right) M_f^2 \, ,
\ee
and from $(i,j)$ components we find
\be
\label{Fbi14}
0 = M_f^2 \left( 2 \dot K + 3 K^2 - 2 LK \right)
+  m^2 M_\mathrm{eff}^2 \left( \frac{a^3c}{b^2} - c^2 \right)
+ \left( %\frac{3}{4} 
\frac{1}{4}{\dot\xi}^2 - \frac{1}{2} U (\xi) c(\tau)^2 \right) 
M_f^2 \, , 
\ee
with $K =\dot b / b$ and $L= \dot c / c$. 
Both Eqs.~(\ref{identity1}) and (\ref{identity2}) yield the identical 
equation:
\be
\label{identity3b}
cH = bK\ \mbox{or}\
\frac{c\dot a}{a} = \dot b\, .
\ee
%%%%
The above equation is the constraint relating the two metrics 
imposed by the equations of motion. 
%%%%
If $\dot a \neq 0$, we obtain $c= a\dot b / \dot a$.
On the other hand, if $\dot a = 0$, we find $\dot b=0$, that is, $a$ and $b$ 
are constant and $c$ can be arbitrary. 

Next we redefine scalars as $\varphi=\varphi(\eta)$ and 
$\xi = \xi (\zeta)$ and identify $\eta$ and $\zeta$ with 
the conformal time $t$ as $\eta=\zeta= \tau $. 
Hence we acquire 
\begin{align}
\label{Fbi19}
\omega(\tau) M_g^2 =&  -4M_g^2 \left( \dot{H}-H^2 \right) -2m^2 M^2_\mathrm{eff}(ab-ac)
\, , \\
\label{Fbi20}
\tilde V (\tau) a(\tau)^2 M_g^2 =& M_g^2 \left (2 \dot{H}+4 H^2 \right ) 
+ m^2 M^2_\mathrm{eff}(6a^2-5ab-ac) \, , \\
\label{Fbi21}
\sigma(\tau) M_f^2 =&  - 4 M_f^2 \left ( \dot{K} - LK  \right )
 - 2m^2 M_\mathrm{eff}^2 \left ( - \frac{c}{b} + 1 \right ) \frac{a^3c}{b^2}
\, , \\
\label{Fbi22}
\tilde U (\tau) c(\tau)^2 M_f^2 =& M_f^2 \left ( 2 \dot{K} + 6 K^2 -2 L K  \right )
+ m^2 M_\mathrm{eff}^2 \left( \frac{a^3c}{b^2} - 2 c^2 + \frac{a^3c^2}{b^3} 
\right) \, , 
\end{align}
with 
\be
\label{Fbi23}
\omega(\eta) = %3 
\varphi'(\eta)^2 \, ,\quad
\tilde V(\eta) = V\left( \varphi\left(\eta\right) \right)\, ,\quad
\sigma(\zeta) = %3 
\xi'(\zeta)^2 \, ,\quad
\tilde U(\zeta) = U \left( \xi \left(\zeta\right) \right) \, .
\ee
Consequently, for arbitrary $a(\tau)$, $b(\tau)$, and $c(\tau)$ 
if we choose $\omega(\tau)$, 
$\tilde V(\tau)$, $\sigma(\tau)$, and $\tilde U(\tau)$ 
to satisfy Eqs.~(\ref{Fbi19})--(\ref{Fbi22}), the cosmological model with 
given evolutions of $a(\tau)$, $b(\tau)$ and $c(\tau)$ can be reconstructed.

%%%%%%%%%%
A reason why we introduced two scalar fields instead of one is that there are 
three degrees of freedom $a$, $b$, and $c$ in metrics (\ref{Fbi10b}) and 
it is not trivial to describe them by using only one scalar field, which 
might not be impossible but we have not succeeded. 
%%%%%%%%%%%%%%%%%%%

\subsection{Conformal description of the accelerating universe \label{subVI1}}

In the following, we use the conformal time. 
We describe how the known cosmologies can be expressed by using the 
conformal time.

The conformally flat FRW universe metric is given by
\be
\label{Fbi31}
ds^2 = a(\tau)^2 \left( - d\tau^2 + \sum_{i=1}^3 \left( dx^i \right)^2
\right) \, .
\ee
In this equation, 
when $a(\tau)^2 = \frac{l^2}{\tau^2}$, 
the metric (\ref{Fbi31}) corresponds to the de Sitter universe, 
which may represent inflation or dark energy in the model under consideration. 
On the other hand if $a(\tau)^2 = \frac{l^{2n}}{\tau^{2n}}$ with $n\neq 1$, by 
redefining the time coordinate as 
\be
\label{Fbi32}
d\tilde t = \pm \frac{l^n}{\tau^n}d\tau\, ,
\ee
i.e., 
\be
\label{Fbi32b}
\tilde t = \pm \frac{l^n}{n-1} \tau^{1-n}\, ,
\ee
the metric (\ref{Fbi31}) can be rewritten as
\be
\label{Fbi33}
ds^2 =  - d{\tilde t}^2 + \left[ \pm (n-1) \frac{\tilde t}{l}
\right]^{- \frac{2n}{1-n}} \sum_{i=1}^3 \left( dx^i \right)^2  \, .
\ee
Equation (\ref{Fbi33}) shows that if $0<n<1$, the metric corresponds 
to the phantom universe \cite{Caldwell:1999ew}, if $n>1$ to 
the quintessence universe, and if $n<0$ to decelerating universe. 
In case of the phantom universe ($0<n<1$), we should choose $+$ sign in $\pm$
of (\ref{Fbi32}) or (\ref{Fbi32b}) and shift $\tilde t$ in (\ref{Fbi33})
as $\tilde t\to \tilde t - t_0$. 
The time $\tilde t=t_0$ corresponds to the Big Rip and 
the present time is 
$\tilde t<t_0$ and the limit of $\tau\to\infty$ is equivalent to 
the infinite past ($\tilde t\to - \infty$). 
In case of the quintessence universe ($n>1$), we may again select $+$ sign in
$\pm$ of (\ref{Fbi32}) or (\ref{Fbi32b}). 
The limit of $\tau\to 0$ corresponds to that of $\tilde t\to + \infty$ and 
that of $\tau\to +\infty$ to that of $\tilde t \to 0$, which may be equivalent 
to the Big Bang. 
In case of the decelerating universe ($n<0$), we may take $-$ sign in $\pm$
of (\ref{Fbi32}) or (\ref{Fbi32b}). 
The limit of $\tau\to 0$ corresponds to that of $\tilde t\to + \infty$ and 
that of $\tau\to +\infty$ to that of $\tilde t \to 0$, which may again be 
considered to be the Big Bang. 
We should also note that in case of the de Sitter universe ($n=1$), 
the limit of $\tau\to 0$ 
corresponds to that of $\tilde t \to + \infty$ and that of $\tau\to \pm \infty$ 
to that of $\tilde t \to - \infty$.

\subsection{Dark energy universe with $a(\tau)=b(\tau)=c(\tau)$ \label{subVI2}}

If the space-time described by the metric $g_{\mu\nu}$ represents 
the universe where we live, the functions $c(\tau)$ and $b(\tau)$ are 
not directly related to the expansion of our universe because the functions 
$c(\tau)$ and $b(\tau)$ correspond to the degrees of freedom in the Einstein frame 
metric $f_{\mu\nu}$. 
Therefore we may choose 
$c(\tau)$ and $b(\tau)$ in the consistent way convenient for the calculation. 
This does not mean $c(\tau)$ and $b(\tau)$ are not relevant for the physics 
besides the expansion of our universe. 
In this section, we simply take $a(\tau)=c(\tau)=b(\tau)$, which satisfy 
the condition (\ref{identity3b}), 
and therefore $H=K=L$. {}From (\ref{Fbi19}) and (\ref{Fbi21}), 
we find $\omega(\tau)=\sigma(\tau)$ and thus $\varphi(\tau)=\xi(\tau)$, and also 
$V(\tau)= U(\tau)$ from (\ref{Fbi20}) and (\ref{Fbi22}). 

By choosing $a(\tau)=c(\tau)=b(\tau)$, 
Eqs.~(\ref{Fbi19}), (\ref{Fbi20}), (\ref{Fbi21}), and (\ref{Fbi22}) are
simplified as
\be
\label{Fbi19BB}
\omega(\tau) = \sigma(\tau) = 4 \left( - \dot H + H^2\right) \, , \quad 
\tilde V (\tau) a(\tau)^2 = \tilde U (\tau) a(\tau)^2 = \left( 2 \dot H + 4 H^2 
\right) \, .
\ee
Let us construct the models, where the scale factor squared is 
given by $a(\tau)^2 = \frac{l^{2n}}{\tau^{2n}}$. 
In this case, we find 
\be
\label{SMT1}
\omega(\tau) = \sigma(\tau) = \frac{4n \left( n - 1 \right)}{\tau^2}\, , 
\quad 
\tilde V (\tau) = \tilde U (\tau) 
= \frac{ \left( 2 n + 4 n^2 \right) l^{2n}}{\tau^{2\left(1 - n \right)} }\, .
\ee
It should be cautioned that if $0<n<1$, $\omega(\tau)$ and $\sigma(\tau)$ become 
negative, and this conflicts with the definition in (\ref{Fbi23}). 
Hence the universe corresponding to the phantom cannot be realized as 
in the standard scalar-tensor model, whose situation is different from 
the case of $F(R)$-bigravity \cite{Nojiri:2012zu} 
(for modified gravity including $F(R)$ gravity and dark energy problem, 
see, e.g.,~\cite{Review-N-O,Clifton:2011jh,Copeland:2006wr,Bamba:2012cp}). 
In case of $n=1$, in which the de Sitter universe is realized, both 
$\omega(\tau)$ and $\sigma(\tau)$ vanish and $\tilde V (\tau)$ and $\tilde U (\tau)$ 
become constants. This is equivalent to the cosmological constant. 

\section{Stability of solutions \label{Sec4}}

As we have shown, a wide class of expansions of the universe can be reproduced 
in the bigravity models coupled to scalar fields. 
The desired solution is, however, only one of the solutions. 
If the solution is not stable under the perturbation, such a solution cannot be realized 
unless we perform very fine-tuning. 
In this section, we study the stability of the solution in the last section. 
For this purpose, we rewrite (\ref{Fbi19})--(\ref{Fbi22}) 
in the following form:
\begin{align}
\label{Fbi19c}
\omega(\eta) {\dot\eta}^2M_g^2 =&  -4M_g^2 \left( \dot{H}-H^2 \right) -2m^2 M^2_\mathrm{eff}
\left(a\left(\tau\right) b\left(\tau\right) - a\left(\tau\right) c\left(\tau\right) \right)
\, , \\
\label{Fbi20c}
\tilde V (\eta) a(\tau)^2 M_g^2 =& M_g^2 \left (2 \dot{H}+4 H^2 \right ) 
+ m^2 M^2_\mathrm{eff}\left( 6a\left(\tau\right)^2 - 5 a\left(\tau\right) 
b\left(\tau\right) - a\left(\tau\right) c\left(\tau\right) \right) \, , \\
\label{Fbi21c}
\sigma(\zeta) {\dot\zeta}^2 M_f^2 =&  - 4 M_f^2 \left ( \dot{K} - LK  \right )
 - 2m^2 M_\mathrm{eff}^2 \left ( - \frac{c\left(\tau\right)}{b\left(\tau\right)} 
+ 1 \right ) \frac{a\left(\tau\right)^3 c\left(\tau\right)}{b\left(\tau\right)^2}
\, , \\
\label{Fbi22c}
\tilde U (\zeta) c(\tau)^2 M_f^2 =& M_f^2 \left ( 2 \dot{K} + 6 K^2 -2 L K  \right )
+ m^2 M_\mathrm{eff}^2 \left( \frac{a\left(\tau\right)^3 c\left(\tau\right)}{b\left(\tau\right)^2} 
 - 2 c\left(\tau\right)^2 + \frac{a\left(\tau\right)^3 c\left(\tau\right)^2}{b\left(\tau\right)^3} 
\right) \, .
\end{align}
On the other hand, the scalar field equations (\ref{scalareq2}) 
can be rewritten to 
\begin{align}
\label{scalareq4}
0& =3 \left(\omega(\eta)\ddot\eta + \frac{\omega'(\eta)}{2}{\dot\eta}^2 
+ 2 H \omega(\eta)\dot \eta \right) 
+ {\tilde V}'(\eta) a^2 \, ,\nn
0& =3 \left(\sigma(\zeta)\ddot\zeta + \frac{\sigma'(\zeta)}{2}{\dot\zeta}^2 
+ \left(3K - L\right) \sigma(\zeta)\dot \zeta \right) 
+ {\tilde U}'(\zeta) a^2 \, .
\end{align}
Equations in (\ref{Fbi19BB}) implies that with a function $f(\tau)$, 
if we choose 
\begin{align}
\label{Fbibi1}
& \omega(\eta) = 4 \left( - f''(\eta) + f'(\eta)^2 \right)\, ,\quad 
\sigma(\zeta) = 4 \left( - f''(\zeta) + f'(\zeta)^2 \right)\, ,\nn
& \tilde V (\eta) = \e^{-2 f(\eta)} \left( 2 f''(\eta) + 4 f'(\eta)^2\right)\, ,\quad 
\tilde U (\zeta) = \e^{-2 f(\zeta)} \left( 2 f''(\zeta) + 4 f'(\zeta)^2\right)\, ,
\end{align}
we find the following solution:
\be
\label{Fbibi2}
a(\tau) = b(\tau) = c(\tau) = \e^{f(\tau)}\, ,\quad \eta = \zeta = \tau  \, .
\ee
We explore the stability of the solution in (\ref{Fbibi2}). 

We may consider the following perturbation: 
\be
\label{Fbibi8b}
H \to H + \delta H\, , \quad K \to K + \delta K\, ,\quad 
a \to a \left( 1 + \delta f_a \right)\, , \quad 
b \to b \left( 1 + \delta f_b \right)\, , \quad 
\eta \to \eta + \delta\eta \, ,\quad \zeta \to \zeta + \delta \zeta\, .
\ee
In what follows, just for simplicity, we take
\be
\label{Fbibi9}
M_f^2 = M_g^2 = \frac{M_\mathrm{eff}^2}{2} = M^2\, .
\ee
Thus we obtain 
\begin{align}
\label{Fbibi19}
& \frac{d}{d\tau} \left( \begin{array}{c} 
\delta \eta \\ \delta \zeta \\ \delta f_a \\ \delta f_b \\ \delta H 
\end{array} \right)
= M \left( \begin{array}{c} 
\delta \eta \\ \delta \zeta \\ \delta f_a \\ \delta f_b \\ \delta H 
\end{array} \right) \, , \nn
& M = \left( \begin{array}{ccccc}
2H & 0 & \frac{C-D}{B} & - \frac{D}{B} & \frac{3}{H B} \left( B - 1 \right) \\
A & E & 2 C - \frac{D}{B} 
& \frac{C + D}{B} - 2C & \frac{3}{H B} \left( B - 1 \right) \\
0 & 0 & 0 & 0 & 1 \\
A H & - A H & 2 H C & - 2 H C & 1 \\
\left( 1 + \frac{D}{3} \right) A \dot H & -\frac{A D \dot H}{3} 
& 2\dot H \left( C - \frac{2 B D}{3} \right) 
&  \frac{4}{3} B D & - 4 H 
\end{array} \right)\, ,
\end{align}
where 
\be
\label{Fbibi20}
A \equiv \frac{\ddot H}{\dot H} + 2H - 4 \frac{H^3}{\dot H}\, ,\quad 
B \equiv 1 - \frac{H^2}{\dot H}\, , \quad 
C \equiv 1 + 2 \frac{H^2}{\dot H}\, ,\quad 
D \equiv \frac{3 m^2 a^2}{\dot H}\, ,\quad 
E \equiv - \frac{\ddot H}{\dot H} - 4 \frac{H^3}{\dot H}\, .
\ee
The derivation of Eqs.~(\ref{Fbibi19}) and (\ref{Fbibi20}) is given in 
Appendix \ref{AA}. 
We should note that we have deleted $\delta K$ in (\ref{Fbibi8b}) 
by using (\ref{Fbibi14}). 

The eigenvalue equation has the following form:
\be
\label{Fbibi21}
0 = \lambda^5 + c_4 \lambda^4 + c_3 \lambda^3 + c_2 \lambda^2 
+ c_1 \lambda + c_0 \, , 
\ee
where $\lambda$ is the eigenvalue of the matrix $M$. 
In order that the solution (\ref{Fbibi2}) could be stable, all the 
eigenvalues should be negative. 
%%%%%
Then all the eigenmodes corresponding to the eigenvalues decrease 
and therefore any perturbation damps. 
%%%%
It requires $c_i > 0$ $\left( i =1,\cdots,4 \right)$. 
Especially $- c_4$ is the trace of the matrix $M$ and we find
\be
\label{Fbibi22}
 - c_4 = - \frac{\ddot H}{\dot H} - 4 H -8 \frac{H^3}{\dot{H}} < 0\, .
\ee
For the power expanding model (\ref{SMT1}), where $H=-n/\tau$, if $\tau>0$, 
Eq.~(\ref{Fbibi22}) leads to
\be
\label{Fbibi23}
4 n^2 + 2 n + 1 < 0 \, .
\ee
Thus there is no real solution for $n$. As a result, there does not exist 
any stable solution for the power expanding model (\ref{SMT1}). 
On the other hand, suppose $\tau <0$, 
Eq.~(\ref{Fbibi22}) yields 
\be
\label{Fbibi24}
4 n^2 + 2 n + 1 > 0 \, , 
\ee
for which there is a possibility that the solution might be stable. 

When $H=-n/\tau$ in (\ref{SMT1}), the matrix $M$ in (\ref{Fbibi19}) has the following form: 
\be
\label{FbibiF1}
M = \left( \begin{array}{ccccc}
 - \frac{2n}{\tau} & 0 & \frac{1 + 2n - D_0 \tau^{-2n+2}}{1 - n} & - \frac{D_0}{1 - n} \tau^{-2n+2} & \frac{3\tau}{1-n} \\
\frac{-2 - 2n + 4n^2}{\tau} & \frac{2+4n^2}{\tau} & 2+4n - \frac{D_0 \tau^{-2n+2}}{1-n} & 
\begin{array}{c} \frac{1 + 2n + D_0 \tau^{-2n +2}}{1-n} \\ - 2 - 4n \end{array}
& \frac{3t}{1-n} \\
0 & 0 & 0 & 0 & 1 \\
 - \frac{n \left( - 2 - 2n + 4n^2 \right)}{\tau^2} & \frac{n \left( - 2 - 2n + 4n^2 \right)}{\tau^2} & 
 - \frac{2n \left( 1 + 2n \right)}{\tau} & \frac{2n \left( 1 + 2n \right)}{\tau} & 1 \\
\left( 1 + \frac{D_0 \tau^{-2n +2}}{3} \right) \frac{ \left(-2 -2n + 4n^2\right) n}{\tau^3} & 
 - \frac{D_0 \left( -2 - 2n + 4n^2 \right) n \tau^{-2n-1}}{3} & 
\begin{array}{c} \frac{2n\left(1+2n\right)}{\tau^2} \\ - \frac{4 \left( 1 - n \right) D_0 \tau^{-2n}}{3} 
\end{array}
& \frac{4\left(1-n\right)D_0 \tau^{-2n+2}}{3} &
\frac{4n}{\tau} 
\end{array} \right)\, , 
\ee
where the scale factor $a(\tau)$ is given by $a=a_0 \tau^{-n}$ and 
$D_0 \equiv \frac{3m^2 a_0^2}{n}$. 
Note that $a_0 = l^n$ in (\ref{SMT1}). 

As an example, we may investigate the case $n=-1/2$. 
In this case, the eigenvalue equation has the following form:
\be
\label{Fbibi26}
0= \lambda  \left( \lambda - \frac{1}{\tau} \right)  \left( \lambda - \frac{3}{\tau} \right) 
\left( 2 D_0 \tau^4 + D_0  \tau^2 - 2 \lambda - \lambda ^2 \tau \right)\, .
\ee
Since there always appear positive eigenvalue, the solution is not stable. 

We redefine
\be
\label{Fbibi27}
\delta f_a = ( 1 - n ) \delta {\tilde f}_a\, ,\quad 
\delta f_b = ( 1 - n ) \delta {\tilde f}_b\, ,\quad 
\delta H = ( 1 - n ) \delta {\tilde H} \, .
\ee
The matrix $M$ in (\ref{FbibiF1}) has the following form:
\be
\label{FbibiF2}
M = \left( \begin{array}{ccccc}
 - \frac{2n}{\tau} & 0 & 1 + 2n - D_0 \tau^{-2n+2} & - D_0 \tau^{-2n+2} & 3t \\
 - \frac{2 ( 1 + 2n ) ( 1 - n)}{\tau} & \frac{2+4n^2}{\tau} & 
\begin{array}{c} 2( 1 + 2n )( 1 - n ) \\ - D_0 \tau^{-2n+2} \end{array} & 
\begin{array}{c} - 1 + 4n^2 \\ + D_0 \tau^{-2n +2} \end{array} & 3t \\
0 & 0 & 0 & 0 & 1 \\
\frac{n \left( 1 + 2n \right)}{\tau^2} & - \frac{n \left( 1 + 2n \right)}{\tau^2} & 
 - \frac{2n \left( 1 + 2n \right)}{\tau} & \frac{2n \left( 1 + 2n \right)}{\tau} & 1 \\
 - \left( 1 + \frac{D_0 \tau^{-2n +2}}{3} \right) \frac{ 1 + 2n }{\tau} & 
\frac{D_0 \left( 1 + 2n \right) n \tau^{-2n - 1}}{3} & 
\frac{2n\left(1+2n\right)}{\tau^2} - \frac{4 \left( 1 - n \right) D_0 \tau^{-2n}}{3}  & 
\frac{4\left(1-n\right)D_0 \tau^{-2n+2}}{3} &
\frac{4n}{\tau} 
\end{array} \right)\, .
\ee
In Section III D, in the part below Eq.~(\ref{SMT1}), 
we have shown that if $0<n<1$, the model is inconsistent. 
As another example, we consider the limit of $n\to 1 + 0$. 
The matrix $M$ in (\ref{FbibiF2}) reduces to 
\be
\label{Fbibi28}
M = \left( \begin{array}{ccccc}
 - \frac{2}{\tau} & 0 & 3 - D_0 & - D_0 & 3t \\
0 & \frac{6}{\tau} & - D_0 & 3+D_0 & 3t \\
0 & 0 & 0 & 0 & 1 \\
\frac{3}{\tau^2} & - \frac{3}{\tau^2} & - \frac{6}{\tau} & \frac{6}{\tau} & 1 \\
- \frac{3 + D_0}{\tau^3} & \frac{D_0}{\tau^3} & \frac{6}{\tau^2} & 0 & \frac{4}{\tau} 
\end{array} \right)\, .
\ee
For this matrix, the eigenvalue equation has the following form:
\begin{align}
\label{Fbibi29}
0=& \lambda^5  - \frac{14}{\tau} \lambda^4 + \left( 6 D_0 + 64 \right) \frac{\lambda^3}{\tau^2}
 - \left( 2 D_0^2 + 66 D_0 + 33 \right) \frac{\lambda^2}{\tau^3} 
+ 3 \left( 8 D_0^2+86 D_0- 63 \right) \frac{\lambda}{\tau^4} \nn
& - 45 \left ( 2 D_0^2 + 3 \right) \frac{1}{\tau^5} \, .
\end{align}
If 
\be
\label{Fbibi30}
\tau <0\ \mbox{and}\ 
D_0 > \frac{\sqrt{2353}-43}{8} = 0.688466417817452 \cdots\, ,
\ee
all the eigenvalues are negative and the system becomes stable. 

In general, the eigenvalue equations (\ref{Fbibi21}) for the matrices 
(\ref{FbibiF1}) and (\ref{FbibiF2}) are rather complicated and the explicit 
forms are given in Appendix \ref{AB}. 
As a result, anyway, we have found a solution which is stable under the perturbation. 
%%%%%%%%%%%
Then we have shown 
that for an arbitrarily given history of the expansion of the universe, we can construct 
a model who has a solution generating the expansion and the solution is stable, that is, 
attractor solution.
%%%%%%%%%%%

\section{Brans-Dicke type model \label{Sec5}}

We introduce a parameter $\epsilon$, which is a positive but sufficiently 
small value ($0 < \epsilon \ll 1$). 
In the previous section, we have found that the model where $n=1+\epsilon$ and 
both $\tau$ and $D_0$ satisfy Eq.~(\ref{Fbibi30}) is stable, 
and that 
the limit $\epsilon\to 0$ $\left( n \to 1 \right)$ corresponds to the de Sitter space. 
In this section, by starting with a model, where $\epsilon>0$ is small enough 
but finite, we construct a model which reproduces an arbitrary expansion 
history of the universe, by using the Brans-Dicke type model. 

We here explore an arbitrary scale factor $a(\tau)$ for $\tau<0$. 
The scale factor corresponding to $n=1+\epsilon$ is given by 
$a_0 \tau^{- 1 - \epsilon}$. 
Hence the metric $g_{\mu\nu}$ corresponding to the scale factor $a(\tau)$ is 
expressed by multiplying the metric $g^\epsilon_{\mu\nu}$ 
corresponding to $n=1+\epsilon$ 
by $a(\tau)^2 a_0^{-2} \tau^{2\left(1 + \epsilon\right)}$. 
Since $\eta= \tau $, we rescale the metric $g_{\mu\nu}$ in the actions (\ref{Fbi1}) 
and (\ref{bimetric2}) as follows
\be
\label{Fbibi31}
g_{\mu\nu} \to a(\eta)^{-2} a_0^2 \eta^{-2\left(1 + \epsilon\right)} g_{\mu\nu} \, .
\ee
By using $\eta$ and $\zeta$, the total action 
$S_\mathrm{total} = S_\mathrm{bi} + S_\varphi + S_\chi$ in 
(\ref{Fbi1}), (\ref{Fbi7b}), and (\ref{bimetric2}) has the following form:
\begin{align}
\label{Fbibi32}
S_\mathrm{total} =&M_g^2\int d^4x\sqrt{-\det g}\,\e^{\Theta(\eta)} R^{(g)}+M_f^2\int d^4x
\sqrt{-\det f}\,R^{(f)} \nn
&+2m^2 M_\mathrm{eff}^2 \int d^4x\sqrt{-\det g}\, \e^{2\Theta(\eta)} 
\left( 3 - \e^{-\frac{\Theta(\eta)}{2}} \tr \sqrt{g^{-1} f}
+ \e^{-2\Theta(\eta)} \det \sqrt{g^{-1} f} \right) \nn
& - M_g^2 \int d^4 x \sqrt{-\det g}
\left\{ \frac{1}{2} 
\e^{\Theta(\eta)} \left( \omega(\eta) - 3 \Theta'(\eta)^2 \right)
g^{\mu\nu} \partial_\mu \eta \partial_\nu \eta
+ \e^{2\Theta(\eta)} \tilde V(\eta) \right\} \nn
& - M_f^2 \int d^4 x \sqrt{-\det f}
\left\{ \frac{1}{2} \sigma(\zeta) f^{\mu\nu} \partial_\mu \zeta \partial_\nu \zeta
+ \tilde U(\zeta) \right\} \, , 
\end{align}
where 
\be
\label{Fbibi33}
\Theta(\eta) \equiv \ln \left( a(\eta)^{-2} a_0^2 \eta^{-2\left(1 + \epsilon\right)} \right)\, .
\ee
Furthermore with (\ref{SMT1}), we have 
\be
\label{Fbibi34}
\omega\left(\eta\right) = \frac{4\left( 1 + \epsilon \right) \epsilon}{\eta^2}\, ,\quad  
\sigma\left(\zeta\right) = \frac{4\left( 1 + \epsilon \right) \epsilon}{\zeta^2}\, ,\quad 
\tilde V \left(\zeta\right) 
= \frac{ 2 \left(1 + \epsilon \right) \left( 3 + 2 \epsilon \right) a_0^2}{\eta^{- 2\epsilon} }\, ,\quad 
\tilde U \left(\zeta\right) 
= \frac{ 2 \left(1 + \epsilon \right) \left( 3 + 2 \epsilon \right) a_0^2}{\zeta^{- 2\epsilon} }
\, .
\ee
We assume that for the Jordan frame of the action (\ref{Fbibi32}), 
the matters do not couple with the scalar fields $\eta$ $\left(\varphi\right)$ 
nor $\zeta$ $\left(\chi\right)$. 
Thus we see that an arbitrary expansion history of the universe can be reproduced 
by the Brans-Dicke type model and the solution is stable by a construction.

\section{Conclusions \label{Sec6}}

In the present paper, 
we have constructed bigravity models coupled with two scalar fields. 
It has been shown that a wide class of the expansion history of the 
universe can be described by a solution of the bigravity model. 
Especially inflation and/or present accelerating expansion can be described 
by this models. 
This situation is very different from the models in the massive gravity, 
where the reference metric is not dynamical. In general, it is very difficult to 
construct a model of the massive gravity, which gives any non-trivial evolution 
of the expansion in the universe. 
The solution obtained in the bigravity model is, however, unstable in general, 
that is, if we add a perturbation 
to the solution, the perturbation grows up. 
Accordingly we have found the conditions for the stability of the solution 
and explicitly constructed a model in which there exists a stable solution.  
The stability can be checked from the eigenvalue equation for the five times five 
matrix in (\ref{Fbibi19}). 
The stable model describes the universe whose expansion is almost that in the 
de Sitter space-time. 
By using the scale transformation of the stable model, 
we construct the Brans-Dicke like model. 
We have shown that the Brans-Dicke type model admits a solution describing 
an arbitrary expanding evolution of the universe. 
The solution is stable, that is, an attractor solution by the construction. 
Therefore even if we started with different initial conditions, which are different 
a little bit with each other, the universe will evolve into the stable solution.  

We should note that the $F(R)$ bigravity models 
in \cite{Nojiri:2012zu,Nojiri:2012re} can be 
rewritten in the scalar-tensor form in (\ref{Fbi1}), 
(\ref{Fbi7b}), and (\ref{bimetric2}) 
by using the scale transformation. 
Therefore we can apply the procedures of the stability analysis in this paper 
to the $F(R)$ bigravity models. 

When we consider the stability, we only consider homogeneous perturbation, which does not depend on the 
spatial coordinates. In case of massive gravity, however, if we consider inhomogeneous perturbation, it has been reported 
that there could appear ghost in inhomogeneous and/or anisotropic background \cite{DeFelice:2012mx}, and there also appear 
superluminal mode in general \cite{Chiang:2012vh}. Furthermore it has been shown that the superluminal mode could break 
causality \cite{Deser:2013eua}. 
Then we need further investigation by using the inhomogeneous perturbation in order to show the consistency in the models 
proposed in this paper. The investigation requires. however, highly non-trivial and complicated calculations. 
Therefore we like to reserve this inhomogeneous perturbation as future works.

\section*{Acknowledgments.}

We are grateful to S.~D.~Odintsov for useful discussions. 
We are also indebted to S. Deser for telling the problem about the superluminality. 
The work is supported by the JSPS Grant-in-Aid for Scientific 
Research (S) \# 22224003 and (C) \# 23540296 (S.N.) 
and that for Young Scientists (B) \# 25800136 (K.B.).  

\appendix

\section{The derivation of Eqs.~(\ref{Fbibi19}) and (\ref{Fbibi20}) \label{AA}}

In this appendix, we derive Eqs.~(\ref{Fbibi19}) and (\ref{Fbibi20}). 

By using (\ref{identity3b}), we have 
\be
\label{Fbibi4}
L = K + \frac{\dot K}{K} - \frac{\dot H}{H}\, .
\ee
Substituting (\ref{identity3b}) and (\ref{Fbibi4}) into 
Eqs.~(\ref{Fbi19c})--(\ref{Fbi22c}), we can eliminate $c$ and $L$ as
\begin{align}
\label{Fbi19d}
\omega(\eta) {\dot\eta}^2M_g^2 =&  -4M_g^2 \left( \dot{H}-H^2 \right) -2m^2 M^2_\mathrm{eff}
a\left(\tau\right) b\left(\tau\right) \left( 1 - \frac{K}{H} \right) \, , \\
\label{Fbi20d}
\tilde V (\eta) a(\tau)^2 M_g^2 =& M_g^2 \left (2 \dot{H}+4 H^2 \right ) 
+ m^2 M^2_\mathrm{eff}\left( 6a\left(\tau\right)^2 - 5 a\left(\tau\right) 
b\left(\tau\right) - \frac{a\left(\tau\right) b\left(\tau\right)  K}{H} \right) \, , \\
\label{Fbi21d}
\sigma(\zeta) {\dot\zeta}^2 M_f^2 =&  - 4 M_f^2 K \left( \frac{\dot{H}}{H} - K  \right)
 - 2m^2 M_\mathrm{eff}^2 \left( 1 - \frac{K}{H} \right) 
\frac{a\left(\tau\right)^3 K}{b\left(\tau\right)H} \, , \\
\label{Fbi22d}
\tilde U (\zeta) b(\tau)^2 M_f^2 =& M_f^2 \left( \frac{2 H \dot H}{K} + 4 H^2 \right)
+ m^2 M_\mathrm{eff}^2 \left( \frac{a\left(\tau\right)^3 H}{b\left(\tau\right) K} 
 - 2 b\left(\tau\right)^2 + \frac{a\left(\tau\right)^3}{b\left(\tau\right)} 
\right) \, .
\end{align}
Furthermore by plugging (\ref{Fbi19d}) into (\ref{Fbi21d}), we find 
\be
\label{Fbibi5}
K - \frac{\sigma(\zeta) {\dot\zeta}^2}{4K} - \frac{m^2 M_\mathrm{eff}^2}{2M_f^2} 
\left( 1 - \frac{K}{H} \right) \frac{a\left(\tau\right)^3}{b\left(\tau\right)H} 
= H - \frac{\omega(\eta) {\dot\eta}^2}{4H} - \frac{m^2 M_\mathrm{eff}^2}{2M_g^2} 
\left( 1 - \frac{K}{H} \right) \frac{a\left(\tau\right) b\left(\tau\right)}{H} \, .
\ee
We also eliminate $\dot H$ from Eqs.~(\ref{Fbi19d}) and (\ref{Fbi20d}) 
and from Eqs.~(\ref{Fbi21d}) and (\ref{Fbi22d}) as follows:
\begin{align}
\label{Fbibi6}
\left( \frac{\omega\left(\eta\right){\dot\eta}^2}{2} + \tilde V \left(\eta\right) 
a\left( t \right)^2 \right) M_g^2 
= 6 M_g^2 H^2 + 6 m^2 M_\mathrm{eff}^2 a \left(\tau\right) 
\left( a\left(\tau\right) - b \left(\tau\right) \right) \, , \\
\label{Fbibi7}
\left( \frac{H^2 \sigma\left(\zeta\right){\dot\zeta}^2}{2K^2} + \tilde U \left(\zeta\right) 
b \left( t \right)^2 \right) M_f^2 
= 6 M_f^2 H^2 - 2 m^2 M_\mathrm{eff}^2 \left( b \left(\tau\right)^2 - \frac{a \left( t \right)^3}{b \left( t \right)} 
\right) \, .
\end{align}
By combining (\ref{Fbibi5}), (\ref{Fbibi6}), and (\ref{Fbibi7}) and deleting $\dot\eta$ and $\dot\zeta$, 
we acquire 
\begin{align}
\label{Fbibi8}
& 0 = 2 (K-H) - \frac{\tilde{U}(\zeta) b(\tau)^2 K}{2H^2}
+ \frac{\tilde{V}(\eta) a(\tau)^2}{2H}
\nonumber \\
& + m^2 M_{\rm eff}^2 \left [
\frac{K}{H^2 M_f^2} \left ( \frac{a(\tau)^3}{b(\tau)}-b(\tau)^2  \right ) - \frac{3}{H M_g^2} 
a(\tau) \left ( a(\tau)-b(\tau) \right )
+ \left ( 1-\frac{K}{H}  \right )
\left ( \frac{a(\tau)^3}{2 M_f^2 b(\tau) H} - \frac{a(\tau) b(\tau)}{2 M_g^2 H}  \right )
\right] \, .
\end{align}
We regard (\ref{Fbi20d}), (\ref{Fbibi6}), (\ref{Fbibi7}), and (\ref{Fbibi8}) 
as independent equations and study the perturbation from the solution 
as in (\ref{Fbibi2}) as in (\ref{Fbibi8b}). We also choose (\ref{Fbibi9}). 
Thus we obtain 
\begin{align}
\label{Fbibi10}
\delta \dot H  =&  \left( - 4 H - \frac{m^2 a^2}{H} \right)  \delta H 
+ \frac{m^2 a^2}{H} \delta K 
+ \left( \ddot H + 2 H \dot H - 4 H^3 \right) \delta\eta \nn
& + \left\{ \left( 2 \dot H + 4 H^2 \right) - 6 m^2 a^2 \right\} \delta f_a 
+ 6 m^2 a^2 \delta f_b  \, , \\
\label{Fbibi12}
2 \left( \dot H - H^2 \right) \delta \dot \eta =& 
4 \left( H \dot H - H^3 \right) \delta \eta + \left( 2 \dot H + 4 H^2 - 6 m^2 a^2 \right) \delta f_a 
 - 6 H \delta H + 6 m^2 a^2 \delta f_b \, , \\
\label{Fbibi13}
2 \left( \dot H - H^2 \right) \delta \dot \zeta =& 
4 \left( H \dot H - H^3 \right) \delta \zeta + \left( 2 \dot H + 4 H^2 + 6 m^2 a^2 \right) \delta f_b 
\nn
& - \frac{2 \left( \dot H - H^2 \right)}{H} \left( \delta H - \delta K \right) 
 - 6 H \delta H - 6 m^2 a^2 \delta f_a  \, , \\
\label{Fbibi14}
 - \frac{\dot{H}}{H^2} (\delta H - \delta K)=& 
\left ( \frac{\ddot{H}}{H} + 2 \dot{H} - 4 H^2  \right ) \left ( \delta
 \eta - \delta \zeta  \right )
+ \left ( 2 \frac{\dot{H}}{H} +4 H  \right ) \left ( \delta f_a - \delta
f_b \right ) \, .
\end{align}
Note that
\begin{align}
\label{Fbibi11}
& \delta V \left( \eta \right) = a \left( t \right)^{-2} \left( 2 \ddot H + 4 H \dot H - 8 H^2 \right) \delta \eta\, ,\quad 
\delta U \left( \zeta \right) = a \left( t \right)^{-2} \left( 2 \ddot H + 4 H \dot H - 8 H^2 \right) \delta \zeta\, ,\nn 
& \delta \omega \left( \eta \right) = 4 \left( - \ddot H + 2 H \dot H \right) \delta \eta \, ,\quad 
\delta \sigma \left( \zeta \right) = 4 \left( - \ddot H + 2 H \dot H \right) \delta \zeta \, .
\end{align}
By using (\ref{Fbibi14}), we may delete $\delta K$ in (\ref{Fbibi10}) and (\ref{Fbibi13}), and eventually we find
\begin{align}
\label{Fbibi16}
\delta \dot{H}
=& -4 H \delta H
+ \left [ \left ( \ddot{H}+2\dot{H}H-4 H^3  \right )
 +m^2 a^2 \left ( \frac{\ddot{H}}{\dot{H}}+2 H-4\frac{H^3}{\dot{H}}  \right )  \right ] \delta \eta
- m^2 a^2  \left ( \frac{ \ddot{H} }{ \dot{H} } + 2 H - 4 \frac{H^3}{\dot{H}} \right ) \delta \zeta 
 \nn
&+ \left [ \left ( 2 \dot{H} + 4 H^2  \right ) -4 m^2 a^2+4 m^2 a^2 \frac{H^2}{\dot{H}}  \right ] \delta f_a
+ \left ( 4 m^2 a^2 - 4 m^2 a^2 \frac{H}{\dot{H}}  \right ) \delta f_b \, , \\
\label{Fbibi17}
\delta \dot{\zeta}
=& \left ( \frac{\ddot{H}}{\dot{H}} +2 H -4 \frac{H^3}{\dot{H}}  \right ) \delta \eta
+ \left ( \frac{\dot{H}+2 H^2 +3 m^2 a^2}{\dot{H}-H^2} -2 -4 \frac{H^2}{\dot{H}}  \right ) \delta f_b
 - \left ( \frac{\ddot{H}}{\dot{H}} +4 \frac{H^3}{\dot{H}}  \right ) \delta \zeta \nn
& + \left ( 2 + 4 \frac{H^2}{\dot{H}}  - \frac{3 m^2 a^2}{\dot{H}-H^2} \right ) \delta f_a
\, .
\end{align}
Since $\delta K= \delta \dot{f}_b$, Eq.~(\ref{Fbibi14}) can be rewritten as 
\be
\label{Fbibi15}
\delta \dot{f}_b = \delta H
+ \left ( \frac{H \ddot{H}}{\dot{H}} + 2 H^2 -4 \frac{H^4}{\dot{H}}  \right )
\left ( \delta \eta - \delta \zeta  \right )
+ \left ( 2 H +4 \frac{H^3}{\dot{H}}  \right ) \left ( \delta f_a-\delta f_b  \right )
\, .
\ee
We may examine the stability by using (\ref{Fbibi12}), (\ref{Fbibi16}), 
(\ref{Fbibi17}), (\ref{Fbibi15}), and the relation 
\be
\label{Fbibi18}
\delta H = \delta \dot{f}_a\, .
\ee

\section{Eigenvalue equations for matrices 
(\ref{FbibiF1}) and (\ref{FbibiF2}) \label{AB}}

In this appendix, we present an explicit forms of the eigenvalue equation 
(\ref{Fbibi21}) for the matrices (\ref{FbibiF1}) and (\ref{FbibiF2}). 

For the matrix (\ref{FbibiF1}), we find 
\begin{align}
\label{ap1}
c_4 =& -\frac{8 n^2+4 n+2}{\tau},\nn
c_3 =& \frac{2}{3} \tau^{-2 (n+1)}\left[2D_0 \tau^2 \left\{5 n^2+n \left(\tau^2+4\right)-\tau^2\right\}
+3 n \left(16n^3+16 n^2+4 n+5\right) \tau^{2 n}\right],\nn
c_2 =& - \frac{2}{3} \tau^{-4 n-3} \left[2D_0^2 n (2 n+1) \tau^4+2D_0 \left\{40 n^4+n^3 \left(8 \tau^2+44\right)
+n^2\left(10-8 \tau^2\right)+n \left(2 \tau^2+5\right)-2 \tau^2\right\} \tau^{2 n+2} \right.\nn
& \left. +3 n \left(32n^4+16 n^3+20 n^2+4 n+3\right) \tau^{4 n}\right],\nn
c_1 =& \frac{4}{3} n \tau^{-4(n+1)} \left[-2D_0 \left\{-32 n^5+8 n^4 \left(4 \tau^2-13\right)
-4 n^3 \left(2\tau^2+13\right)-n^2 \left(24 \tau^2+29\right)+n \left(\tau^2-14\right)-\tau^2-3\right\} \tau^{2n+2} \right. \nn
& \left. +4 n \tau^4 (2D_0 n+D_0)^2+3 \left(32 n^4+8 n^3-2 n^2-n-1\right)\tau^{4 n}\right],\nn
c_0 =& \frac{4}{3} n^2 (2 n+1) \tau^{-4 n-5} \left[ 2D_0^2\tau^4 \left\{-8 n^3+8 n^2 \left(\tau^2-2\right)-8 n \tau^2-3\right\}
\right. \nn
& \left. + 4D_0 \left(4n^2-5 n+1\right) \left(n-\tau^2\right) \tau^{2 n+2} 
+3 \left(16 n^3+8 n^2+2 n+1\right)\tau^{4 n} 
\right] \, .
\end{align}
and for and (\ref{FbibiF2}), 
\begin{align}
\label{ap2}
c_4 =& - \frac{8 n^2+4 n+2}{\tau},\nn
c_3 =& - \frac{1}{3} \tau^{-2 (n+1)}\left[D_0 \tau^2 \left\{-2 n^2-n \left(10 \tau^2+7\right)+\tau^2\right\}
-3 \tau^{2 n}\left\{24 n^4+28 n^3-2 n^2+n \left(6 \tau^2+5\right)+3 \tau^2\right\}\right],\nn
c_2 =& - \frac{1}{3} \tau^{-4 n-3} \left[2D_0^2 (2 n+1) \tau^6+D_0 \left\{16 n^4+8n^3 \left(10 \tau^2+9\right)
-10 n^2+n \left(34 \tau^2+3\right)+3 \tau^2\right\} \tau^{2 n+2} \right. \nn
& \left. +3\left\{48 n^5-72 n^4+4 n^3 \left(12 \tau^2-5\right)+2 n^2 \left(16 \tau^2-9\right)
+2 n\left(7 \tau^2-2\right)+5 \tau^2\right\} \tau^{4 n}\right],\nn
c_1 =& - \frac{1}{3} \tau^{-4(n+1)} \left[2D_0^2 (2 n+1) \tau^4 \left\{6 n^3
+n^2 \left(1-14 \tau^2\right)+n\left(2-5 \tau^2\right)-2 \tau^2\right\}\right.\nn
& -D_0 \left\{-48 n^6+16 n^5 \left(3\tau^2+22\right)+n^4 \left(96 \tau^2-88\right)
+8 n^3 \left(40 \tau^2-13\right) \right. \nn
& \left. +3 n^2\left(48 \tau^2-5\right)+n \left(57 \tau^2+2\right)+10 \tau^2\right\} \tau^{2 n+2} \nn
& \left. +3 (2 n+1)\left\{144 n^5-8 n^4 \left(9 \tau^2+1\right)-4 n^3 \left(5 \tau^2-7\right)-2 n^2 \left(3\tau^2+1\right)
-3 n \tau^2+2 \tau^2\right\} \tau^{4 n}\right],\nn
c_0 =& - \frac{1}{3} n (2 n+1)\tau^{-6 n-5} \left[2D_0^3 \tau^6 \left\{-2 n^2+n \left(2\tau^2-1\right)+\tau^2\right\} \right. \nn
& +2D_0^2 \left\{-8 n^4+8 n^3 \left(\tau^2+3\right)+2 n^2\left(12 \tau^2-7\right)+n \left(8 \tau^2-2\right)+5 \tau^2\right\} \tau^{2 n+4} \nn
& +D_0\left\{80 n^5-16 n^4 \left(5 \tau^2+13\right)+32 n^3 \left(5 \tau^2+1\right)-4 n^2\left(8 \tau^2-3\right)
+3 n \left(8 \tau^2+1\right)+9 \tau^2\right\} \tau^{4 n+2} \nn
& \left. +9 \left(8n^3+4 n^2+2 n+1\right) \left(4 n^2-2 n \tau^2-\tau^2\right) \tau^{6 n}\right] \, .
\end{align}

\end{document}